\documentclass[review]{elsarticle}

\usepackage{epsfig}
\usepackage{amsmath}
\usepackage{graphics}
\usepackage{placeins}
\usepackage{color}
\usepackage{lineno,hyperref}
\modulolinenumbers[250]
\begin{document}

\begin{frontmatter}

\title{Effect of the chromo-electromagnetic field fluctuations on heavy quark  propagation at the LHC energies}


\author[mymainaddress]{Ashik Ikbal Sheikh}
\author[mymainaddress]{Zubayer Ahammed* }
\author[mysecondaryaddress]{Prashant Shukla}
\author[mytertiaryaddress]{Munshi G.  Mustafa}

\cortext[mycorrespondingauthor]{za@vecc.gov.in}

\address[mymainaddress]{Variable Energy Cyclotron Centre, HBNI, 1/AF Bidhannagar, Kolkata 700 064, India}
\address[mysecondaryaddress]{Nuclear Physics Division, Bhabha Atomic Research Centre, HBNI, Mumbai 400085, India}
\address[mytertiaryaddress]{Theory Division, Saha Institute of Nuclear Physics, HBNI, 1/AF Bidhannagar, Kolkata 700 064, India }

\begin{abstract}
We consider the effect of the chromo-electromagnetic 
field fluctuations in addition to  the collisional as well as the radiative energy loss  suffered by heavy quarks 
while propagating through the hot and densed  deconfined medium of quarks and gluons created in relativistic heavy ion collisions. 
The chromo-electromagnetic field fluctuations play an important role as it leads to an energy gain of heavy quarks of all momentum, 
significantly effective at the lower momentum region. With this, we have computed, for the first time, the nuclear 
modification factor ($R_{AA}$) of heavy mesons,
 {\textit viz.}, {\it D}-mesons and {\it B}-mesons and compared  with the those experimental measurements 
  in  $Pb-Pb$ collisions at  $\sqrt{s_{NN}} = 2.76 \, $TeV and  $\sqrt{s_{NN}} = 5.02 \, $TeV by the CMS and ALICE experiments  
 at the LHC.  Our results are found to be in very good agreement with those available data measured by CMS and ALICE  experiments.
\end{abstract}

\begin{keyword}
\texttt{Quark-Gluon Plasma, Chromo-electromagnetic Field Fuctuations, Heavy Quarks, Energy Loss, Energy Gain}
\end{keyword}

\end{frontmatter}


\section{Introduction}
\label{intro}
The main goal of relativistic heavy-ion collisions at Relativistic Heavy Ion Collider (RHIC) at BNL and Large Hadron Collider (LHC) at CERN 
is to produce a hot and dense deconfined state of QCD matter, so called quark-gluon plasma (QGP). It is believed that this new deconfined state 
of matter has been formed during relativistic heavy-ion collisions at RHIC\cite{RHIC} and LHC\cite{LHC}. One of the features of this deconfined 
plasma created in heavy ion collisions is the suppression of high energy hadrons compare to the case of $p-p$ collisions, called jet quenching. 
This jet quenching is caused due to the energy loss of initial hard partons via collisional and radiative energy loss inside the deconfined medium. 
It was anticipated first by Bjorken\cite{Bjorken} as a crucial probe of this deconfined medium. 
In view of this the energy loss suffered by highly energetic partons, both light and heavy quarks, in the deconfined QCD medium  is a field 
immense interest because it unravels the dynamic properties of the medium.  
Particularly, heavy quarks by their nature in this deconfined medium remains an active and important field of research\cite{TG,BT,MG,mgm05,dokshit01,dead,Fochler:2010wn,alam10,Abir10,das10,Fochler:2008ts,Gossiaux10,horowitz10,wicks07,jamil10,armesto1,armesto2,yd,Uphoff11a,Uphoff11b,PP,Fl,AJMS,KPV,KPV1}.

 Heavy quarks are mostly produced in early stage of the heavy ion collisions from the initial fusion of 
 partons 
 They may also be produced in the QGP, 
 if initial temperature of QGP is high enough than the mass of the heavy quarks. However, no heavy quarks are produced at the 
 latter stage and none in the hadronic matters. 
 Hence, the total number of heavy quarks  becomes frozen at the very early stage in the history of the collisions, 
 which makes them a good probe of the QGP. These heavy quarks immediately after their production  will propagate through the dense medium and 
 will start losing energy during their path of travel. This energy loss suffered by the heavy quarks are reflected 
 in the transverse momentum spectra and nuclear modification factor of heavy mesons. 
Heavy quarks lose energy in two different fashions in the QGP: one is caused by elastic collisons with the light partons 
 of the thermal background (QGP) and the other one is by radiating gluons, {\it viz.}, bremsstrahlung process due to the deceleration
 of the charge particles. 
 
 The energy loss in the QGP  are usually obtained by treating the medium in an average manner and the fluctuations are ignored.
 Since QGP is a statistical ensemble of mobile coloured charge particles, which  could also be characterised by 
 omnipresent stochastic fluctuations. This microscopic fluctuations generally couple with the external perturbations and affect 
 the response of the medium.  The effect of electromagnetic field fluctuations during the passage of charged particles though a non-relativistic 
 classical plasma has been  calculated by several authors in the literature\cite{Gasirowicz,Sitenko,Akhiezer,Kalman,Thompson,Ichimaru}. 
 On the other hand the  effect of chromo-electromagnetic  fluctuations in the QGP leads to an energy gain 
 of heavy quarks of all momentum and significantly at the lower momentum ones\cite{Fl}. 
 This is because the moving parton in the QGP encounters a statistical change  in the energy  due to  the fluctuations of the 
 chromo-electromagnetic fields as well as the velocity of the particle  under the influence of this field.
The effect of such fluctuation was not considered in earlier literature for studying the hadron spectra in the perspective of heavy ion collisions. 
 
 In this Letter,  we investigate for the first time the effect of the chromo-electromagnetic field  fluctuations leading to energy gain
of heavy quarks in addition to both the collisional and the radiative energy loss  on 
 the nuclear modification  factor for $D$ and $B$ mesons and compared with the measurements of both ALICE and CMS experiments in $Pb-Pb$ 
 collisions at $\sqrt{s_{NN}} = 2.76$ TeV and CMS experiment at $\sqrt{s_{NN}} = 5.02$ TeV. We found that the chromo-electromagnetic field fluctuations 
 play an important role on the propagation of  the heavy quark jets  in a QGP  vis-a-vis the nuclear modification factor of heavy flavoured hadrons.
 It is interesting to note that only collisional or radiative or both energy loss can not explain the data satisfactorily. If the energy gain due to fluctuations 
 is included along with the collisional and  radiative energy loss, then the data can be explained very satisfactorily from low to moderately high value 
 of transverse momentum. 
 
 The paper is organised in as follows: In sec.\ref{sec2} we brifely outline the basic setup containing, heavy quark production and fragmentation, 
 models for both collisional and radiative 
 energy loss, and energy gain due to field fluctuations, medium evolution and initial conditions etc,  for the purpose. Here we consider the collisional energy loss 
 of heavy quarks by Peigne and Pashier (PP) formalism\cite{PP} and the radiative energy loss by Abir, Jamil, Mustafa and Srivastava (AJMS) 
 formalism\cite{AJMS} along with the energy gain 
 due to the chromo-electric field fluctuations  prescription by Chakraborty, Mustafa and Thoma (CMT) in Ref.\cite{Fl}.  In sec.\ref{sec3} we discuss our results 
 and  a conclusion in sec.\ref{sec4}.

\section{Methodology}
\label{sec2}

\subsection{Heavy quark production and fragmentation}
 The heavy quarks in $p-p$ collisions are mainly produced by fusion of gluons or light quarks\cite{Combridge}. Their production cross section has been obtained to
 next-to-leading order (NLO ) accuracy with CT10 parton distribution function\cite{CT10} for $p$-$p$ collisions. For heavy ion collisions, the shadowing 
 effect is taken into account by using the NLO parameters of EPS09\cite{EPS09} nuclear parton distribution function. The same 
 set of parameters as that of Nelson et. al.\cite{Nelson} have been used. For fragmentation of $c$ quarks into $D$-mesons and $b$ quarks into
  $B$-mesons, the Peterson fragmentation function\cite{Peterson} with parameters $\epsilon_{c} = 0.016$ for $c$ quarks and $\epsilon_{b} = 0.0012$ 
  for $b$ quarks have been used. For other  details of the production and fragmentation of heavy quarks we refer the readers to Refs.\cite{AJMS,KPV,KPV1}
    
\subsection{Medium Evolution and initial condition}
 As the heavy quarks lose energy during their passage through the QGP medium, it is important to figure out the path length
  it is traversing inside the medium. We consider a heavy quark, which is being produced at a point ($r$,$\phi$) in heavy ion 
  collisions and propagates at an angle $\phi$ with respect to $\hat{r}$ in the transverse plane. So, 
  the path length $L$ covered by the heavy quark inside the medium is given by~\cite{Muller}:
\begin{equation}
 L(r,\phi) = \sqrt{R^{2}-r^{2}\sin^{2}{\phi}} - r\cos{\phi}.
\end{equation}
where  $R$ is the radius of the colliding nuclei.  The average distance travelled by the heavy quark inside the plasma is
\begin{align}
\label{eq2}
 \langle L \rangle &= \frac{\int\limits_{0}^{R}rdr\int\limits_{0}^{2\pi}L(r,\phi)T_{AA}(r,b=0)d\phi}{\int\limits_{0}^{R}rdr\int\limits_{0}^{2\pi}T_{AA}(r,b=0)d\phi} , 
\end{align}
where $T_{AA}(r,b=0)$ is the nuclear overlap function. We estimate $\langle L \rangle = 6.14 fm$ for central $Pb-Pb$ collisions. 
The effective path length of heavy quark of transverse mass $m_T$ and transverse momentum $p_T$ in the QGP of life time $\tau_f$ is obtained as,
\begin{align}
 L_{\mbox{eff}} &= \mbox{min}[L, \frac{p_T}{m_T} \times \tau_f].
\end{align}
 We consider the medium evolution as per the isentropic cylindrical expansion as discussed in Ref.\cite{Expansion}. 
 The equation of state is obtained by Lattice QCD along with hadronic resonance in order to calculate temperature as a function 
 of proper time\cite{EOS}. We calculate the heavy quark energy loss over QGP life time and finally averaged over the 
 temperature evolution. The initial conditions used for the hydrodynamic medium evolution are similar to the Ref~\cite{KPV1}. 
We consider the  initial time $\tau_0 = 0.3$ fm and  freeze-out  time $\tau_f = 6$ fm. At $\sqrt{s_{NN}} = 2.76$ TeV, impact 
parameter $b = 9.68$, number of participents $N_{part} = 113$, $dN/d\eta = 363$, $L = 4.16$ fm and $T_0 = 0.436$ GeV for (0 - 100)\% 
centrality and $b = 3.44$, $N_{part} = 356$, $dN/d\eta = 1449$, $L = 5.73$ fm and $T_0 = 0.467$ GeV for (0 - 10)\% centrality 
have been taken. For $\sqrt{s_{NN}} = 5.02$ TeV, we take $b = 9.65$, $N_{part} = 114$, $dN/d\eta = 463$, $L = 4.18$ fm 
and $T_0 = 0.469$ GeV for (0 - 100)\% centrality and $b = 3.34$, $N_{part} = 359$, $dN/d\eta = 1749$, $L = 5.74$ fm 
and $T_0 = 0.508$ GeV for (0 - 10)\% centrality.

\subsection{Collisional Energy Loss: Peigne and Peshier (PP) Formalism}
 One of the important mechanism in which heavy quarks may lose energy inside the QGP is by collisions. The calculation of collisional energy loss
  per unit length $dE/dx$ has been reported by in the past by several authors\cite{TG,BT,Alex}. The most detailed calculation of $dE/dx$ was made 
  by Brateen and Thoma \cite{BT} which was based on their previous QED calculation of $dE/dx$ for muon \cite{Muon}. This calculation of 
 Brateen and Thoma for $dE/dx$ is based on an assumption that the momentum exchange in elastic collisions, $q \ll E$, which is not appropriate in the
 domain $E \gg M^{2}/T$, where $M$ is the mass of the heavy quark and $T$ is the temperature of the medium. 
The improved differential energy loss expression, valid for $E \gg M^{2}/T$,  is given by Peigne and Pashier~\cite{PP} as
\begin{align}
\frac{dE}{dx} &= \frac{4\pi\alpha_{s}^{2}T^{2}}{3}\left[\left(1+\frac{n_f}{6}\right)\log\left(\frac{ET}{\mu_{g}^{2}}\right)+\frac{2}{9}\log\left(\frac{ET}{M^{2}}\right)+c(n_f)\right],
\end{align}
 where, $\mu_{g}^{2} = 4\pi\alpha_{s}T^{2}\left(1+{n_f}/{6}\right)$ is the square of Debye screening mass, $n_f = 3$, is the number 
 of active quark flavours and $c(n_f) \approx 0.146n_f+0.05$ and $\alpha_{s} = 0.3$, is the strong constant.

\subsection{Radiative Energy Loss: Abir, Jamil, Mustafa and Srivastava (AJMS) Formalism}
 The most important and dominant way of energy loss from a fast partons inside the QGP is due to gluon radiation. The first attempt to estimate 
 the radiative energy loss was made in Ref.\cite{MG}. Later many authors\cite{dokshit01,dead,wicks07,armesto2,B.Z,W.C,Vitev} also estimated the energy loss
 with various ingredients and kinematical conditions. In Refs.\cite{dokshit01,dead} the soft gluon emission was estimated  which was found to suppress compared to
 the light quarks  due to the mass effect, known as dead cone effect.  The radiative energy loss induced by the medium due to the dead cone effect was limited 
 only to the forward direction.  In Ref.\cite{Abir10} by relaxing some of the constraints imposed in Refs.\cite{dokshit01,dead}, 
 e.g., the gluon emission angle and the scaled mass of the heavy quark with its energy, a generalised  dead cone  was obtained which led to a very compact 
 expression for the  gluon emission  probability  off a heavy quark. Based on the generalised dead cone approach and the gluon emission
 probability,   AJMS~\cite{AJMS} computed the heavy quark radiative energy loss\footnote{Later a kinematical correction was made in Ref.\cite{KPV}.} as
\begin{align}
\frac{dE}{dx} &= 24\alpha_{s}^{3}\rho_{QGP}\frac{1}{\mu_g}\left(1-\beta_1\right)\left(\sqrt{\frac{1}{(1-\beta_1)}\log(\frac{1}{\beta_1})}-1\right)\mathcal F(\delta),
\end{align}
 with 
\begin{align}
\mathcal F(\delta) &= 2\delta - \frac{1}{2}\log\left(\frac{1+\frac{M^2}{s}e^{2\delta}}{1+\frac{M^2}{s}e^{-2\delta}}\right)-\left(\frac{\frac{M^2}{s}\sinh{(2\delta)}}
{1+2\frac{M^2}{s}\cosh{(2\delta)}+\frac{M^4}{s^2}} \right),
\end{align}
where
\begin{align}
\delta &= \frac{1}{2}\log\left[\frac{1}{(1-\beta_1)}\log\left (\frac{1}{\beta_1}\right )\left(1+\sqrt{1-\frac{(1-\beta_1)}{\log\frac{1}{\beta_1}}}\right)^2\right].
\end{align}
and $\rho_{QGP}$  is the density of the QGP medium which acts as a background containing the target partons. If $\rho_q$ and $\rho_g$ are the density 
of quarks and gluons respectively in the medium, then the $\rho_{QGP}$ is given by
\begin{align}
\rho_{QGP} &= \rho_q+\frac{9}{4}\rho_g,
\end{align}
\begin{align}
\beta_1 &= \frac{\mu_{g}^{2}}{CET},
\end{align}
\begin{align}
 C &= \frac{3}{2}-\frac{M^2}{4ET}+\frac{M^4}{48E^2T^2\beta_0}\log\left(\frac{M^2+6ET(1+\beta_0)}{M^2+6ET(1-\beta_0)}\right),
\end{align}
\begin{align} 
\beta_0 &= \sqrt{1-\frac{M^2}{E^2}}.
\end{align}
\subsection{Energy gain by chromo-electromagnetic fields fluctuations: Chakraborty, Mustafa and Thoma (CMT) Formalism}
The energy loss calculations both collisional and radiative of heavy quarks in the QGP were obtained by treating the QGP medium 
without considering microscopic fluctuations. However, QGP being the statistical system, it is characterised by stochastic chromo-electromagnetic 
field fluctuations. Since  the energy loss is of topical interest for the phenomenology of  heavy quark  jet quenching in hot and dense medium.
A quantitatively estimate of the effect of the  microscopic electromagnetic fluctuations on the propagation a heavy quark was done using 
semiclassical approximation\footnote{This approximation has been shown to be equivalent to the Hard Thermal Loop approximation 
which is based on the weak coupling limit\cite{TG,BT} and also referred to Ref.\cite{Fl} for other details.} by CMT in Ref.\cite{Fl}.  
This  was found to led  an energy gain 
of the heavy quark caused due to the  statistical change in the energy of the moving parton in the QGP due to the fluctuations 
of the chromo-electromagnetic fields as well as the velocity of the particle under the influence of this field.  The leading-log (LL) contribution 
of the energy gain was obtained\cite{Fl} as
\begin{align} 
\left(\frac{dE}{dx}\right)_{\mbox{fl}}^{\mbox{LL}} &= 2\pi C_F\alpha_{s}^{2}\left(1+\frac{n_f}{6}\right)\frac{T^3}{Ev^2}\ln{\frac{1+v}{1-v}}\ln{\frac{k_{\mbox{max}}}{k_{\mbox{min}}}},
\end{align}
where $k_{\mbox{min}} = \mu_g = $ Debye mass and $k_{\mbox{max}} = \mbox{min}\left[E,\frac{2q(E+p)}{\sqrt{M^2+2q(E+p)}}\right]$ with $q \sim T$ is the typical momentum 
of the thermal partons.  One can physically interpret this energy gain of a heavy quark that absorbs gluons during its propagation.
\begin{figure}[!h]
  \centering
  \begin{minipage}[b]{0.4\textwidth}
    \includegraphics[width=\textwidth]{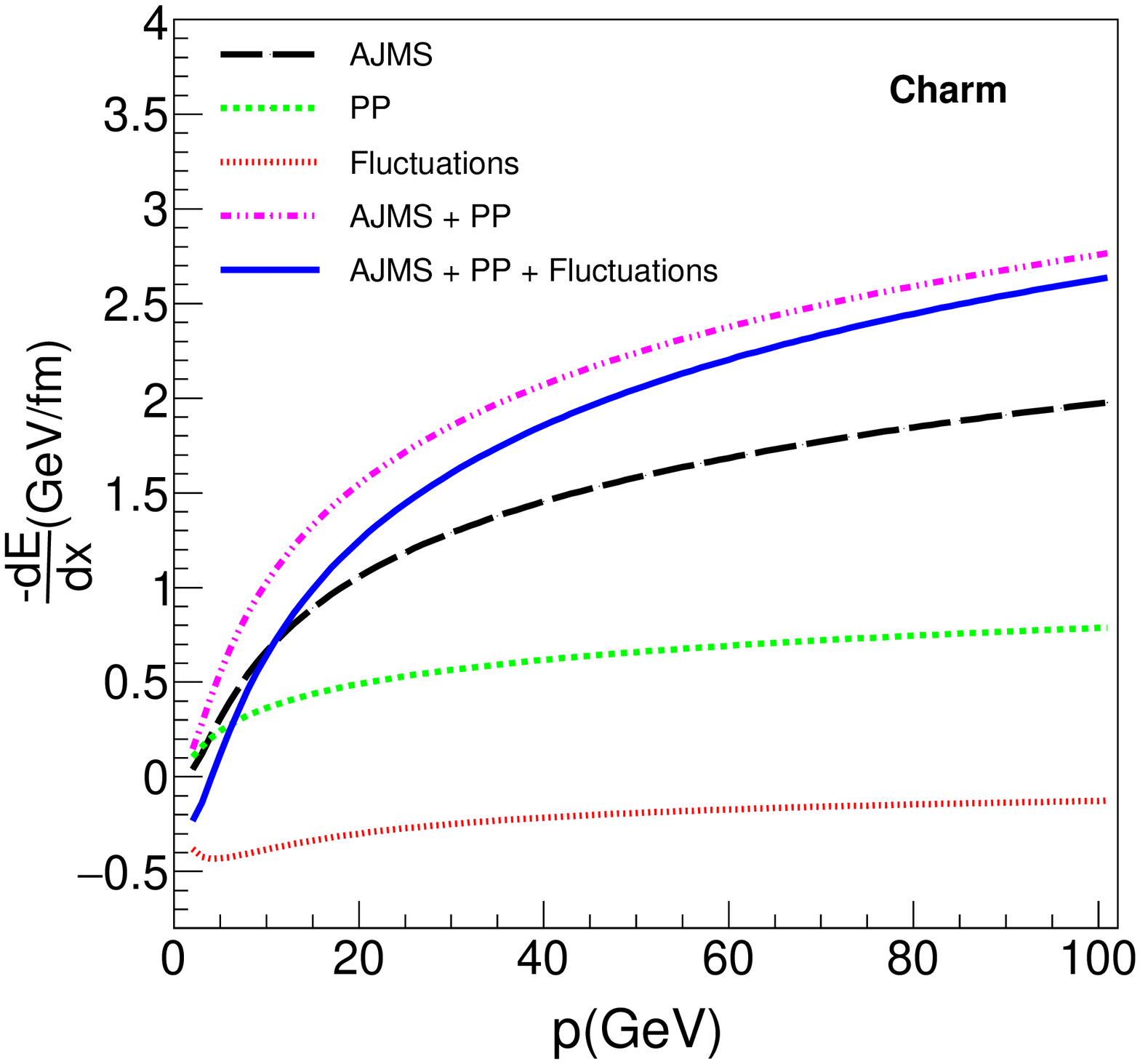}
    \caption{The energy loss of a charm quark inside QGP medium as a function of its momentum, 
    obtained using PP\cite{PP}, AJMS\cite{AJMS} and Fluctuations\cite{Fl}.}
    \label{dedx_c}
  \end{minipage}
  \hfill
  \begin{minipage}[b]{0.4\textwidth}
    \includegraphics[width=\textwidth]{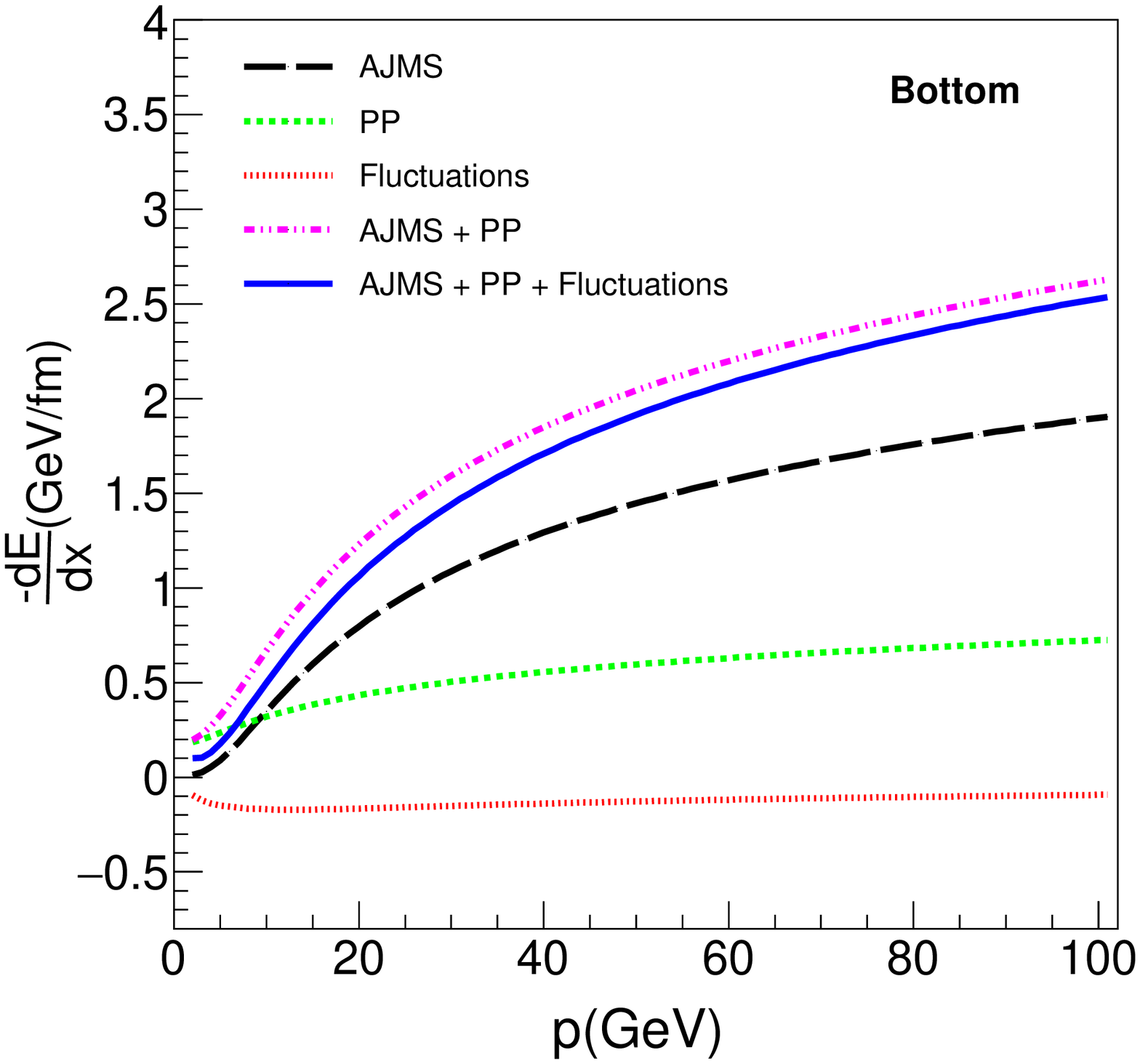}
    \caption{The energy loss of a bottom quark inside QGP medium as a function of its momentum, obtained using PP\cite{PP}, AJMS\cite{AJMS} 
    and Fluctuations\cite{Fl}.}
     \label{dedx_b}
  \end{minipage}
\end{figure}
\FloatBarrier

In Fig.\ref{dedx_c} and Fig.\ref{dedx_b}, we show the various contributions to the energy loss of charm and bottom quarks respectively. 
 The effect of field fluctuations is also shown here. Our choices of parameters are: $n_f=2$, $\alpha_s = 0.3$, charm quark mass, $M_c= 1.25$ GeV 
 and  bottom quark mass, $M_b= 4.2$ GeV.  As seen that the AJMS radiative energy loss  always dominates the PP collisional energy loss 
 for charm quark whereas 
 for bottom quark, the PP collisional energy loss  dominates the AJMS radiative energy loss upto  momentum ~$10$ GeV, beyond  which
 the AJMS radiative energy takes over. The differential energy loss  ($-dE/dx$) is negative due to the field fluctuations which 
 implies energy gain. The energy gain due to the field fluctuations is found to be significant in the momenta range ($4 -20$) GeV. 
 This energy gain is relatively more for charm quark compared to bottom quark.  
   
\begin{figure}[!h]
  \centering
  \begin{minipage}[b]{0.4\textwidth}
    \includegraphics[width=\textwidth]{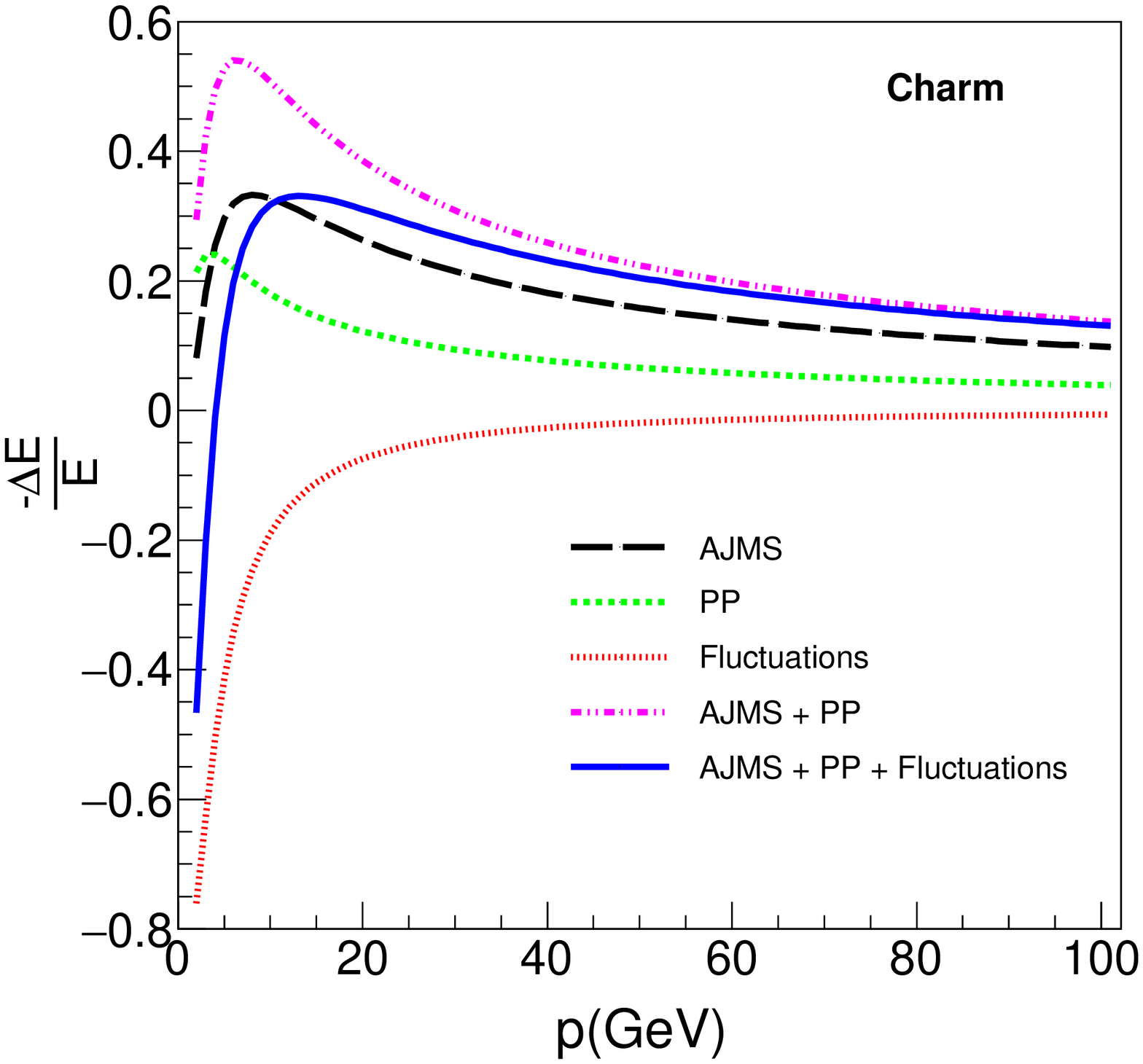}
    \caption{Fractional energy loss of charm quark inside QGP due to fluctuations, collisions (PP) and radiations (AJMS) 
    with its momentum. The path length considered is $L=5$ fm.}
      \label{de_c}
  \end{minipage}
  \hfill
  \begin{minipage}[b]{0.4\textwidth}
    \includegraphics[width=\textwidth]{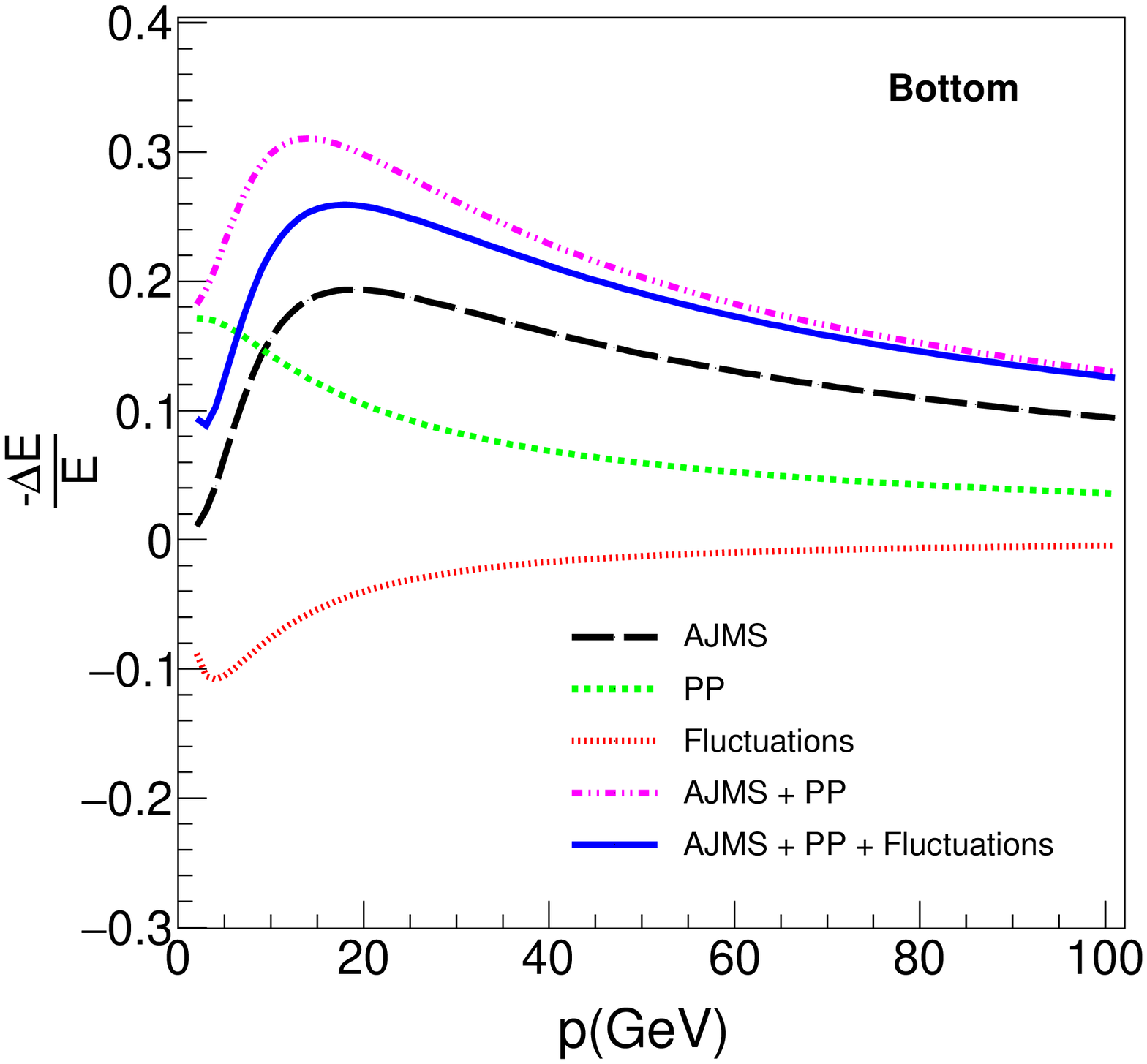}
    \caption{Fractional energy loss of bottom quark inside QGP due to fluctuations, collisions (PP) and radiations (AJMS) 
    with its momentum. The path length considered is $L=5$ fm.}
      \label{de_b}
  \end{minipage}
\end{figure}
\FloatBarrier
 
  Fig.\ref{de_c} and Fig.\ref{de_b} display the fractional energy loss from collisional and radiative process, and  also 
 the energy gain due to the field fluctuations for charm and bottom quarks, respectively. It is clear that the energy gain for 
 heavy quarks is relatively more at the lower momentum  region ($4-40$ MeV) than that in very higher 
 momentum ($>40$ MeV)  region.  The reason is that the field fluctuations and thus the energy gain become 
 substantial in  the low velocity limit.  Because of this the field fluctuations, the total energy loss of a heavy quark gets 
 reduced  up to a very moderately high values of momentum beyond which its contribution gradually diminishes. 
 The relative importance of it will be very re
levant for LHC energies as we would see below. 
 
\section{Results and Discussions}
\label{sec3}

\begin{figure}[!ht]
  \centering
  \begin{minipage}[b]{0.4\textwidth}
    \includegraphics[width=\textwidth]{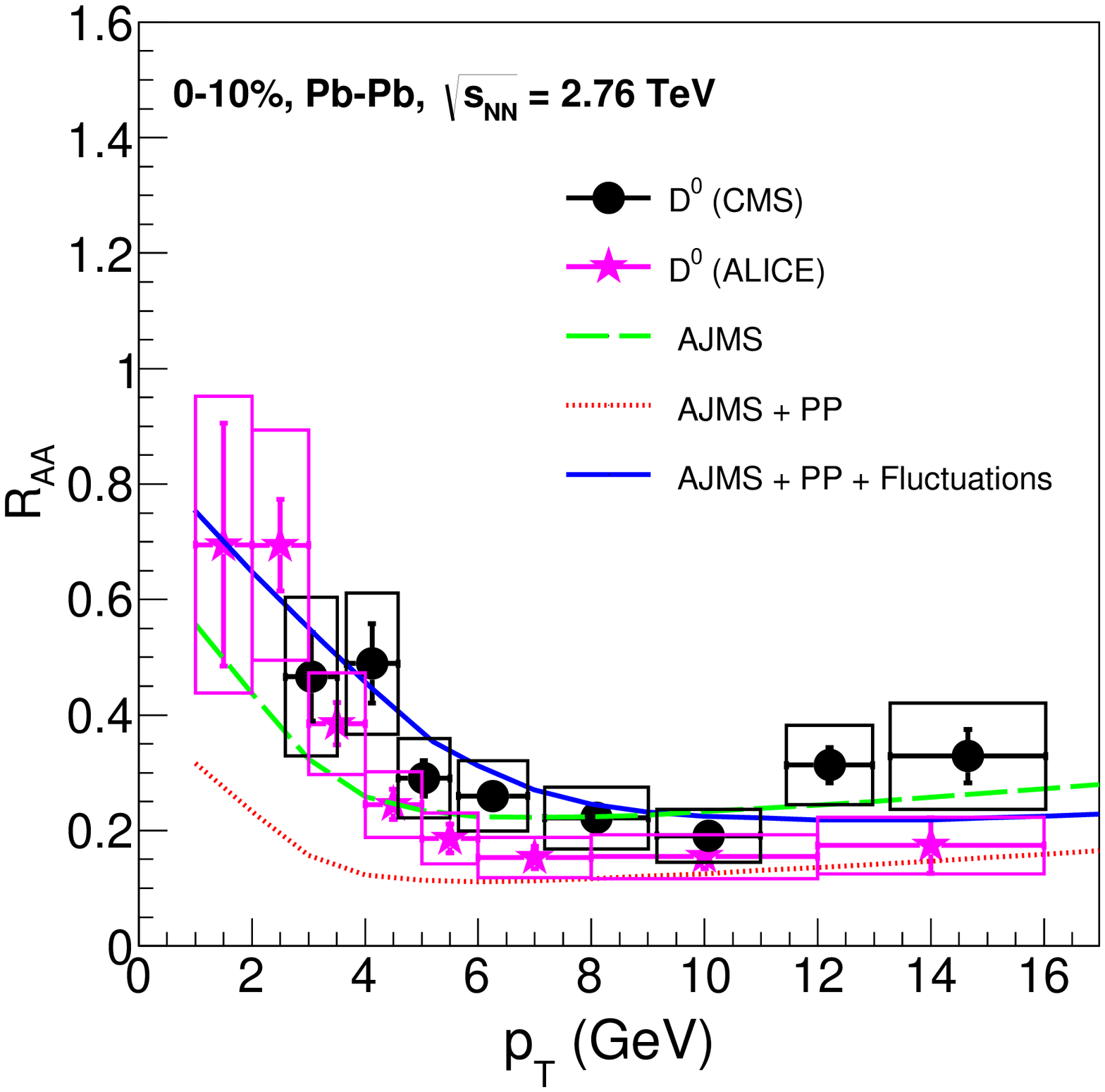}
    \caption{Nuclear modification factor $R_{AA}$ of $D^0$-meson with collisional (PP) and radiative (AJMS) energy loss along with 
    the effect of fluctuations as a function of transverse momentum $p_T$ for $(0-10)\%$ centrality at $Pb-Pb$ collisions 
    at $\sqrt{s_{NN}} = 2.76$ TeV. The data for $D^0$-meson are taken from the measurement of ALICE~\cite{ALICE_D} 
    and CMS experiment~\cite{CMS_D_2TeV}.}
     \label{raa_d0_10}
  \end{minipage}
  \hfill
  \begin{minipage}[b]{0.4\textwidth}
    \includegraphics[width=\textwidth]{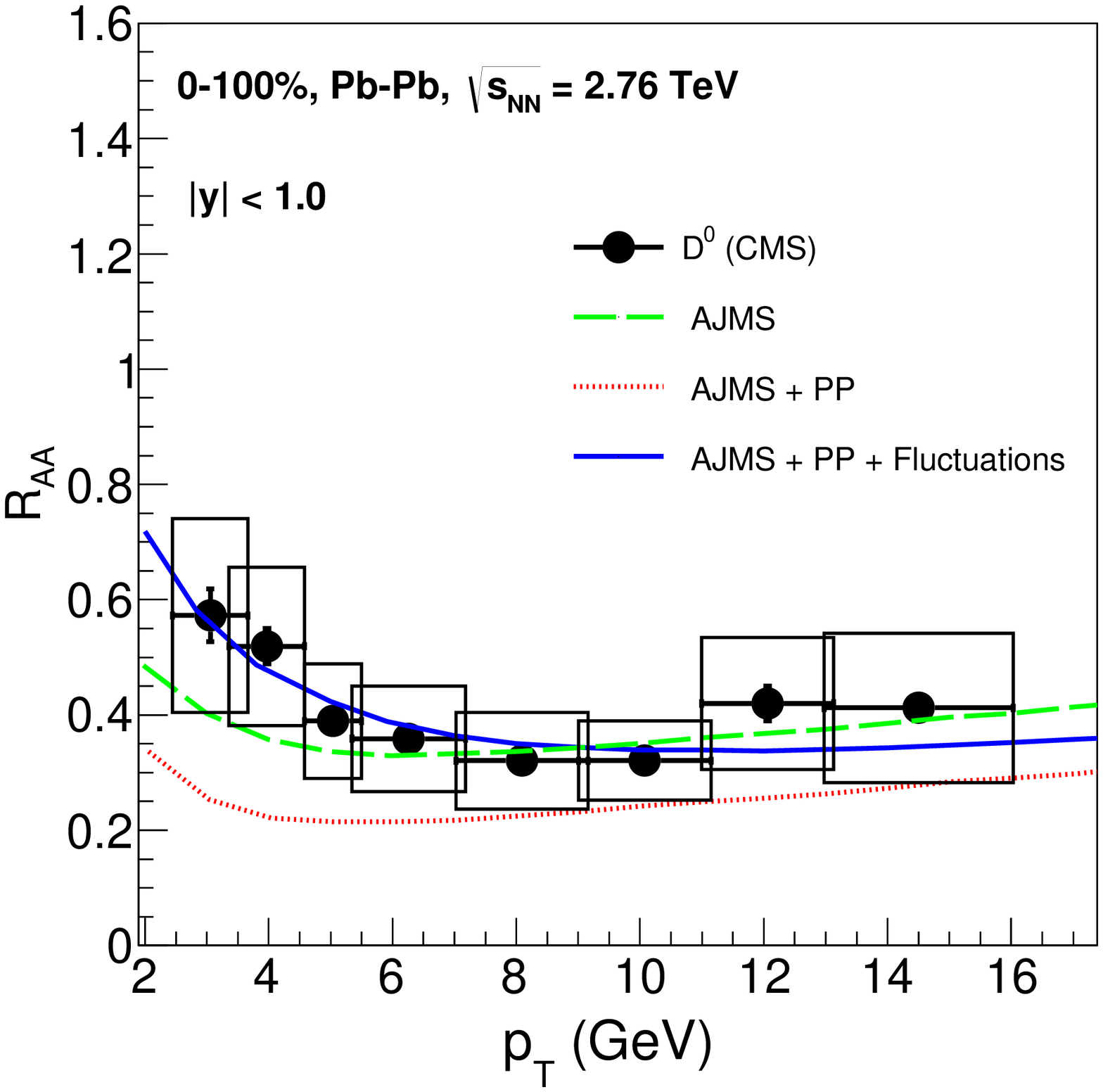}
    \caption{Nuclear modification factor $R_{AA}$ of $D^0$-meson with collisional(PP) and radiative(AJMS) energy loss along 
    with the effect of fluctuations as a function of transverse momentum $p_T$ for $(0-100)\%$ centrality in $Pb-Pb$ collisions
     at $\sqrt{s_{NN}} = 2.76$ TeV. The data for $D^0$-meson are taken from the measurement of CMS experiment~\cite{CMS_D_2TeV}.}
      \label{raa_d0_100}
  \end{minipage}
\end{figure}

   In Fig.\ref{raa_d0_10} and Fig.\ref{raa_d0_100} we have displayed the nuclear modification factor, $R_{AA}$, for $ D^0 $-meson in 
 ($0-10$)$\% $ and  ($0-100$)$\% $ centrality, respectively,  in $Pb-Pb$ collisions, considering both collisional and radiative energy loss
 along with the energy gain due to the  field  fluctuations and compared with ALICE~\cite{ALICE_D} and CMS data~\cite{CMS_D_2TeV}. 
 We observe that only the radiative  energy loss (AJMS) \, or \, the 
\FloatBarrier

\noindent collisional energy loss (PP) along with the radiative energy loss can not explain
  the data properly.  As seen that the radiative along with the collisional energy loss overestimate the data in entire 
  $p_T$ range whereas the radiative one 
  alone can describe the date for the transverse momentum $p_T>10$ GeV because the radiative energy loss is the dominant 
  one compared to that of the collisional one.  When the energy gain due to the field fluctuations were included in addition the both 
  collisional and radiative losses the measured data can be nicely described in the entire $p_T$ range.  The field fluctuations is found to play an
  important role in the phenomenology of the heavy quark jet quenching observed in the LHC energies.
  
\begin{figure}[!ht]
  \centering
  \begin{minipage}[b]{0.4\textwidth}
    \includegraphics[width=\textwidth]{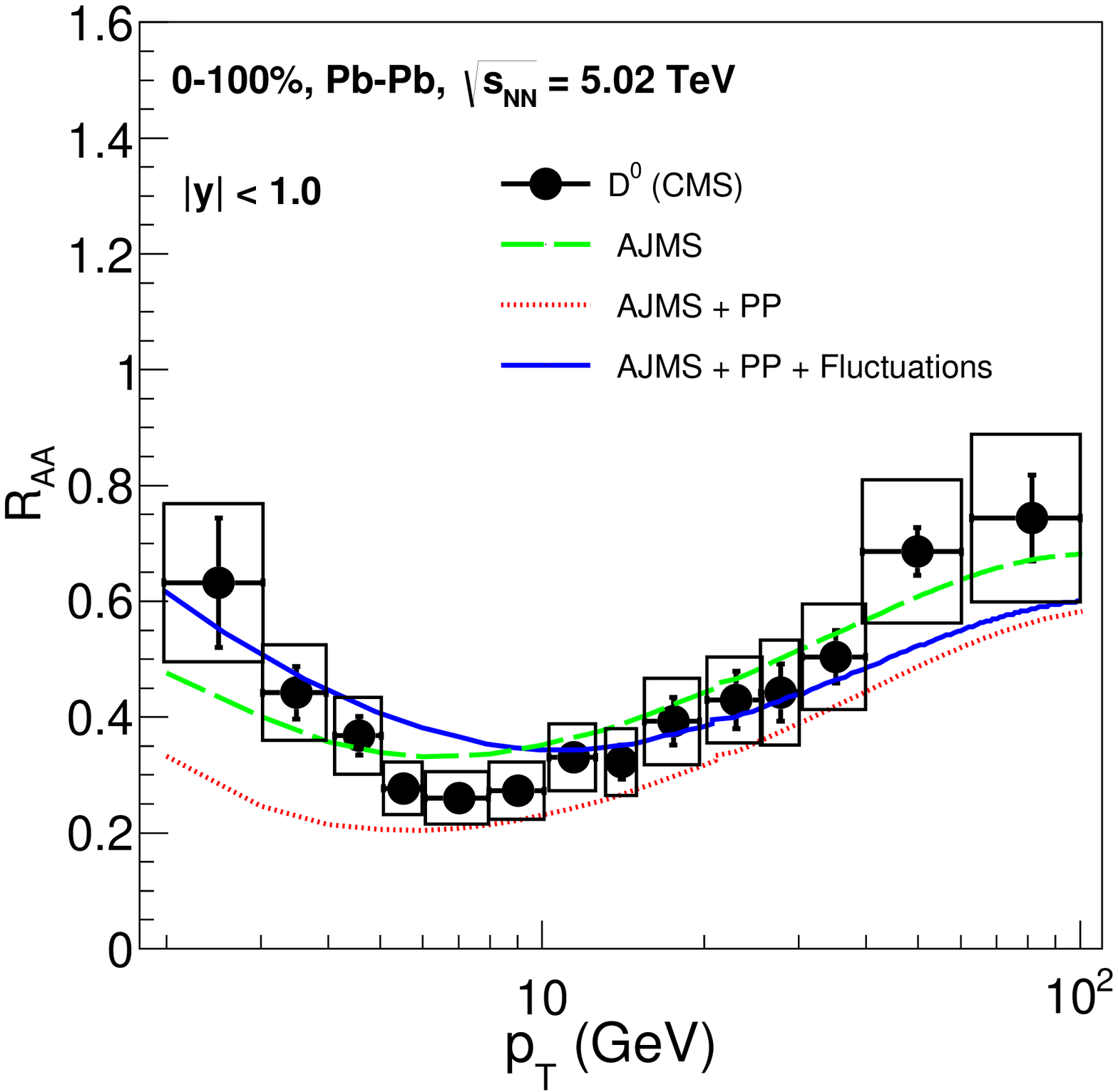}
    \caption{Nuclear modification factor $R_{AA}$ of $D^0$-meson with collisional (PP) and radiative (AJMS) energy loss along with the effect of 
    fluctuations as a function of transverse momentum $p_T$ for $(0-100)\%$ centrality in $Pb-Pb$ collisions at $\sqrt{s_{NN}} = 5.02$ TeV. 
    The data for $D^0$-meson are taken from the measurement of CMS experiment~\cite{CMS_D_5TeV}.}
     \label{raa_d0_c_502}
  \end{minipage}
  \hfill
  \begin{minipage}[b]{0.4\textwidth}
    \includegraphics[width=\textwidth]{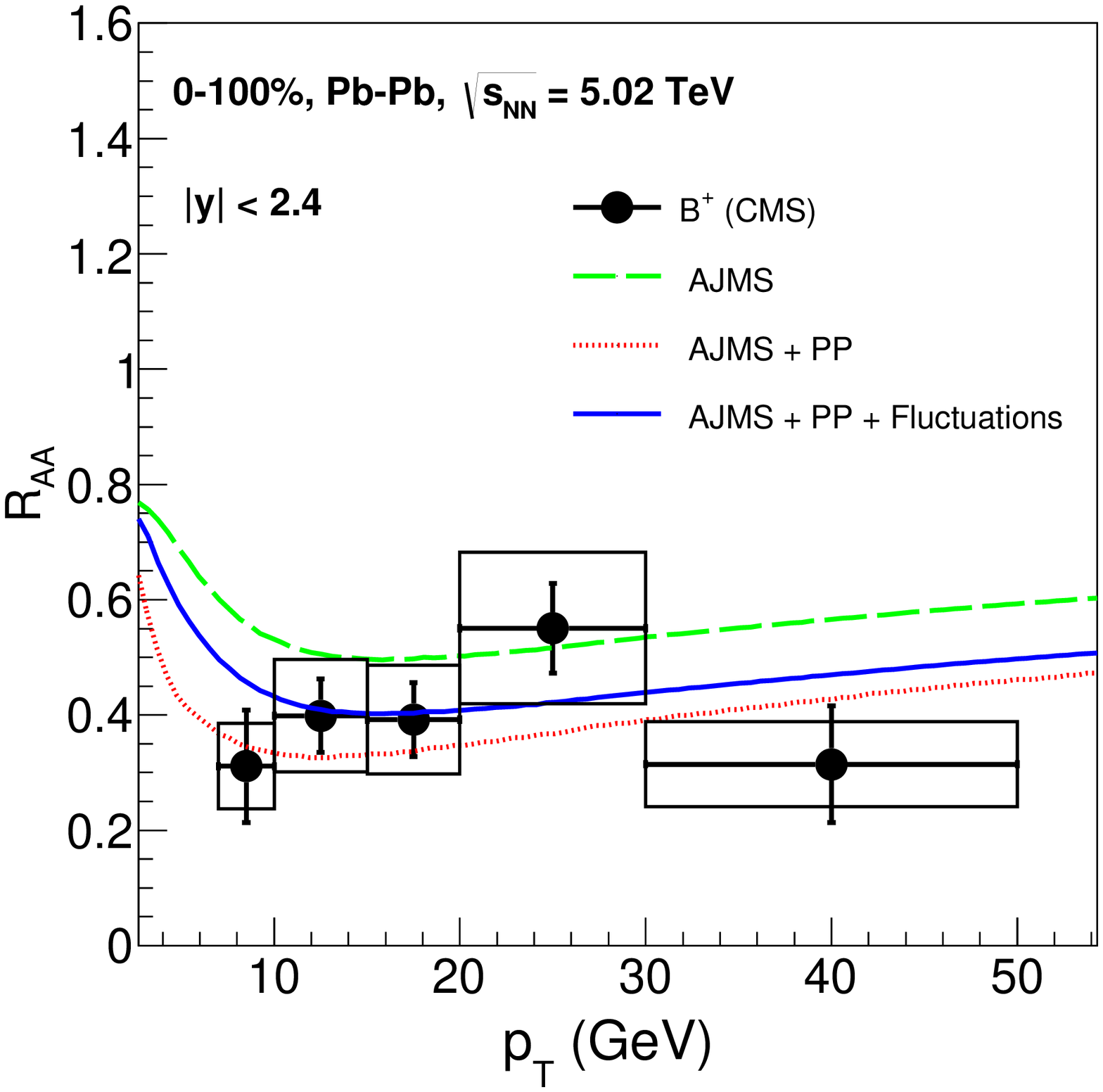}
    \caption{Nuclear modification factor $R_{AA}$ of $B^+$-meson with collisional (PP) and radiative (AJMS) energy loss along with 
    the effect of fluctuations as a function of transverse momentum $p_T$ ffor $(0-100)\%$ centrality in $Pb-Pb$ collisions 
    at $\sqrt{s_{NN}} = 5.02$ TeV. The data for $B^+$-meson are taken from the measurement of CMS experiment~\cite{CMS_B_5TeV}.}
      \label{raa_d0_b_502}
  \end{minipage}
\end{figure}

 Fig. \ref{raa_d0_c_502} displays the nuclear modification factor $R_{AA}$ of $D^0$-meson as a function of transverse momentum $p_T$, 
 obtained using collisional (PP), radiative(AJMS) energy loss and field fluctuations in $Pb-Pb$ collisions at $\sqrt{s_{NN}} = 5.02$ TeV. 
 The experimental data is obtained from the CMS collaboration~\cite{CMS_D_5TeV}.  Again the radiative energy loss (AJMS) alone can describe 
 the data above transverse momentum 10 GeV but the  $R_{AA}$ spectra  in full $p_T$ range can be described when the effect of fluctuations 
 is taken into consideration.

\FloatBarrier

 In Fig \ref{raa_d0_b_502} the nuclear modification factor $R_{AA}$, for $B^+$-meson in $Pb-Pb$ collisions at $\sqrt{s_{NN}} = 5.02$ TeV is 
 displayed considering both collisional and radiative energy loss along with the effect of the field fluctuations and compared with CMS data~\cite{CMS_B_5TeV}. 
 The radiative energy loss itself  produces a small suppression but when the  collisional one is added generates more suppression than the  
 measured CMS data. When the energy gain due to field fluctuation is taken into account in addition to both radiative and collisional losses, 
 the suppression is  found to be very closer to the measured data  within their uncertainties. 
 
\section{Conclusion}
\label{sec4}
 The  energy loss encountered by an energetic parton in a QGP medium reveals the dynamical properties 
 of that medium in view of jet quenching of high energy partons. This is  usually reflected in the transverse momentum spectra and 
nuclear modification factor of mesons which are measured in heavy ion experiments.  
For the phenomenology of the  heavy quarks jet quenching  the field fluctuations in the QGP medium
were not considered in the literature before. In this article, for the first time, we have considered the propagation of 
high energy heavy quarks  by including the 
energy gain due to field fluctuations along with the energy loss  caused by the collisions and gluon radiations inside the QGP medium. 
The nuclear modification factors $R_{AA}$ for $D$-mesons and $B$-mesons in $Pb-Pb$ collisions  
at $\sqrt{s_{NN}} = 2.76$ TeV and $\sqrt{s_{NN}} = 5.02$ TeV  are calculated by including 
the both energy losses and the field fluctuations effect. We note that the radiative energy loss  alone can describe 
the $D$-mesons suppressions at higher transverse momentum. Nevertheless, the nuclear modification factors for both $D$ and $B$ mesons 
are found to agree quite well with those data in the entire $p_T$ range measured by CMS and ALICE experiments at LHC energies, 
if the energy gain
due the field fluctuations are taken into account  in addition to the collisional and radiative loss in the medium. The effect of field fluctuations 
in hot and dense QGP medium is found to play an important role in the propagation of heavy quarks 
also in describing the experimental data for heavy quarks quenching. 

\section{Acknowledgement}
 ZA would like to acknowledge the discussions with Kapil Saraswat and Subhasis Chattopadhyay.

\end{document}